# Data-driven Decision Support on Student's Behavior using Fuzzy-Based Approach

Jerry M. Lumasag[1], Hidear Talirongan[2], Florence Jean B. Talirongan[3] & Charies L. Labanza[4]

[1-4]*Misamis Misamis University, Ozamiz City, Misamis Occidental, Philippines.*
*Email: jerry.lumasag@mu.edu.ph[1], hidear@mu.edu.ph[2], badilles.fj@gmail.com[3] & labanza.c@gmail.com[4]*

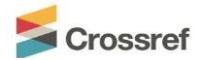

Country: Philippines

## ABSTRACT

Monitoring of students' behavior in school needs further consideration in order to lessen the number of casualties in every term. The study designs a data-driven decision support on students' behavior utilizing Fuzzy-Based Approach. The study successfully produces common behavioral problems of the student and able to give interventions for the improvement of students' behavior. Student behavioral problems identified were absenteeism, tardiness and poor academic performance.

## 1. Introduction

High-availability data systems both in research and decision support communities gain an increase of demand because of the growth of internet application that is used for everyday activity (Mohamed at al., 2020; Marinakis et al., 2018). People look for data to strengthen decision-making, but high quality information may not always be available (Price & Shanks, 2016). Yet such data are effective implementation. The use of high quality data facilitates management lessens reliance on unreliable information, and guarantee that data are accessible for decision making. Decision Support System (DSS) are effective tools for gathering high quality data (Aiello et al., 2018). DSS is a type of information system which focuses on manipulating and modeling data in order to enhance the procedures in decision-making. It enables the users to go through a large amount of data and gather information that can be used to solve problems and make better decision. It can be categorized into five types. These are communication driven, model driven, knowledge driven, document driven and data driven (Ada & Ghaffarzadeh, 2015; Belkadi et. al, 2020). From the five categories, data-driven DSS is the most common decision support systems in the expanded DSS framework. The major differences are given by the way the data and knowledge are stores and processed (Omidipoor et. al, 2020; Brown-Liburd et. al, 2015).

Schools nowadays are adapting DSS that mainly concerns on the academic performance of the students, but student academic behavior are not being taken into consideration (Masmiquel et al., 2017). Students go to schools with a goal of pursuing and finishing studies through employment (Norman et. al, 2015; Owens, 2017). In order to achieve the student's aim to finish schooling, regular attendance and being attentive in the class are the techniques they applied. During classroom hours, it was observed that some students were identified for absenteeism, tardiness, poor academic performance and creating disturbances of misconduct and these students could be referred to the Guidance Center for counseling. However, there are reasons why these students are being referred for the said demeanor. No note-taking, inattentive during class and irregular study habit are students' common reason during interview with the guidance counselors. Schools nowadays are integrating DSS concerning student's academic performance to improve student achievement particularly in difficult subjects which supports in the conduct of making decisions. It was built and investigated for approximately 40 years using multiple methods and frameworks (Migueis et al., 2018; Kombo, 2018).

A local study of Solpico et al. (2015) on web-based decision system for Philippines lakes used UAV imaging to gain better understanding of the lake and to make better decisions. Moreover, a web-based visual DSS helped the organization to successfully turn letter shop environment spanning transaction (Krishnaiyer & Chen, 2017). Archimedes IndiGo is a sophisticated DSS that analyzes patient-specific health information including lab results, diagnoses, medications, weight, blood pressure, and risk factors extracted from electronic sources.

The DSS used advanced algorithms to create individualized guidelines, displayed in graphical representation (Bellows et al., 2014). Brynjolfsson & McElheran (2019) examined whether firms that use data-driven decision making or DDD showed higher performance. Automation for the selection of envelope and structural systems for educational systems needs the application of Decision Support System (Alshamrani & Alshibani, 2020). Datnow and Hubbard (2016) stated that data-driven decision making is very essential in the educational planning of teachers and even the management position.

In Misamis University, there are data that are available in the Guidance and Testing Center but are not fully utilized. With respect to the needs of the deans and guidance counselor, existing reports lack details to help them provide better decisions concerning student referrals and generating reports take a lot of time and effort to finish on the current practice that they are adopting.

The study aimed to visualize the academic behavior and mental ability test results of all referred students of each college in Misamis University. This will be realized with the application of data-driven decision support system.

## 2. System Architecture

Figure 1 presents the research framework of the study comprising the component in coming up with the visualization that include input, process and output.

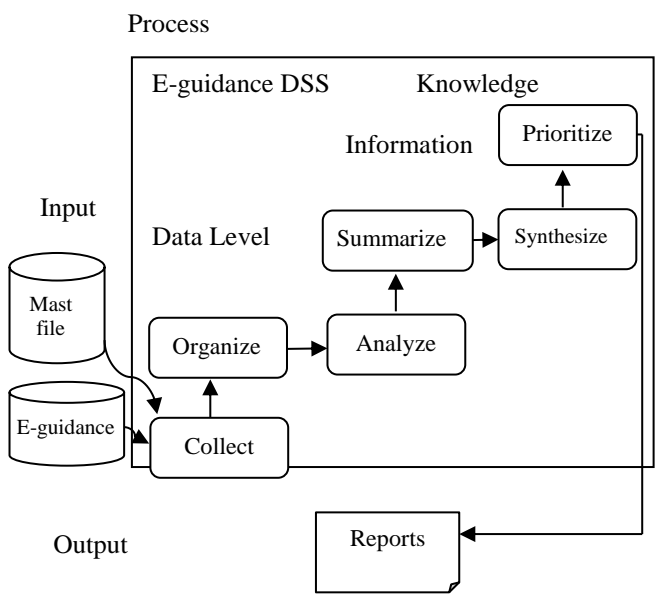

**Figure 1:** Research Framework

The study adapted the framework of Mandinach (2015). There were two different sources, the mastfile and e-Guidance where relevant data were collected and organized from. The system then analyzed and summarized these data to produce information significant in deciding for solution to the students' problems. The analyzed and synthesized information will then be used as an output of the system.

*Data Level*

Collection and organization of data were the tasks performed during this level. The data were collected from two data sources – masterfile and e-guidance databases. These data were organized into DSS database, a new database designed to store the collected data.

*Information Level*

The data from the DSS database were analyzed for informational purposes. Data analysis was dependent on the inquiry of the user. Furthermore, the system summarized of all accumulated information to produce concise information before it was transformed into usable knowledge.

*Knowledge Level*

In turning this information into knowledge, the system synthesized and prioritized the available information. The system generated reports provided the knowledge to the user.

**3. Methodology**

The study used the Mamdani-style inference model for the process states especially in the decision making pertaining to the student behavior and determining proper intervention. Tables 1, 2 and 3 present the input variables number of poor academic performance, tardiness and absenteeism while table 4 shows the output variable on student's intervention. Figures 2, 3, 4 and 5 show the corresponding graphs of the tables. The data came from the referred data on students in the Guidance Office of Misamis University, Ozamiz City, Philippines who happened to be the expert. To visualize further, figure 6 shows the fuzzy inference system of the study.

**Table 1:** Input Variable: Poor Academic Performance

| Interpretation | Poor Academic Performance |
|---|---|
| Low | 0 – 3 |
| Medium | 1 – 5 |
| High | 2 – 7 |

**Table 2:** Input Variable: Tardiness

| Interpretation | Tardiness |
|---|---|
| Low | 0 – 4 |
| Medium | 3 – 8 |
| High | 6 – 12 |

Table 3: Input Variable: Absenteeism

| Interpretation | Absenteeism |
|---|---|
| Low | 0 – 3 |
| Medium | 1 – 5 |
| High | 2 – 7 |

Table 4: Output Variable: Intervention

| Interpretation | Intervention |
|---|---|
| Workshop & Counseling | 0.0 – 2.0 |
| Tutoring & Advisor | 2.1 – 4.0 |
| Lighter load & Study more | 4.1 – 6.0 |

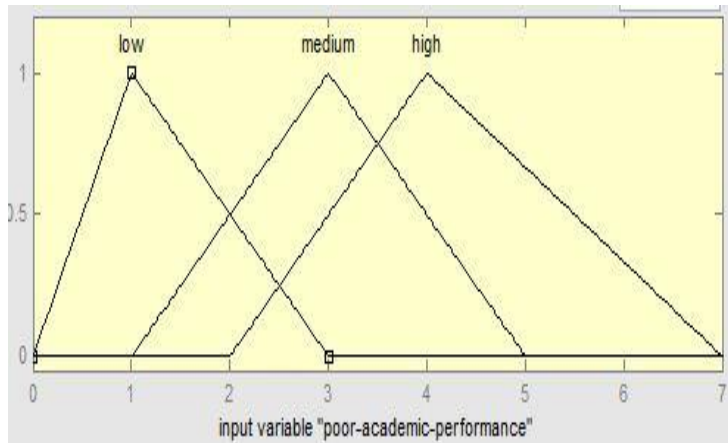

**Figure 2:** Graph of the Input Variable, Poor Academic Performance

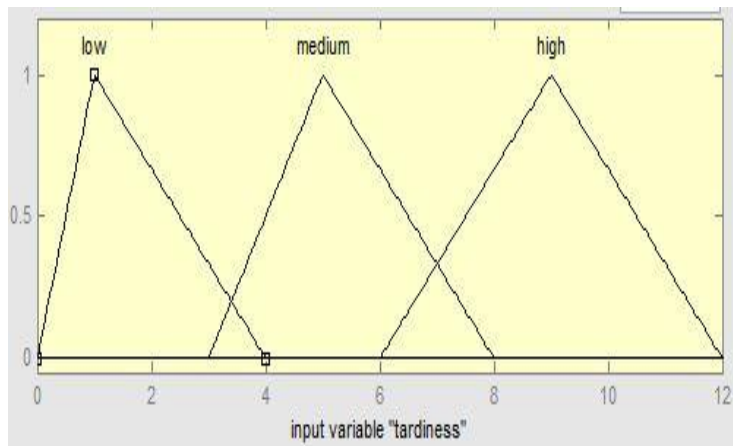

**Figure 3:** Graph of the Input Variable, Tardiness

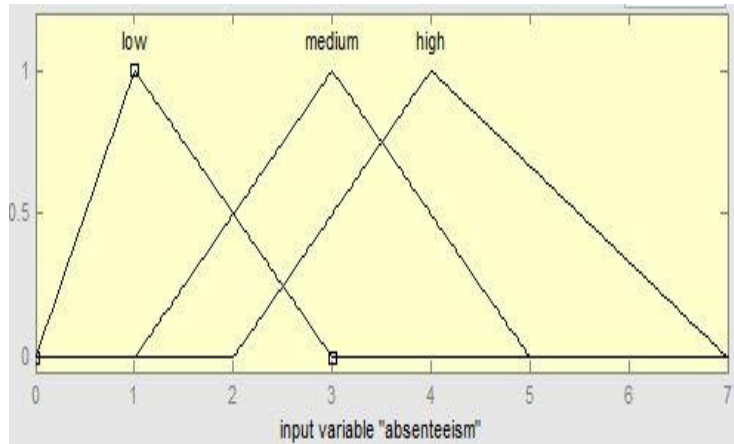

**Figure 4:** Graph of the Input Variable, Absenteeism

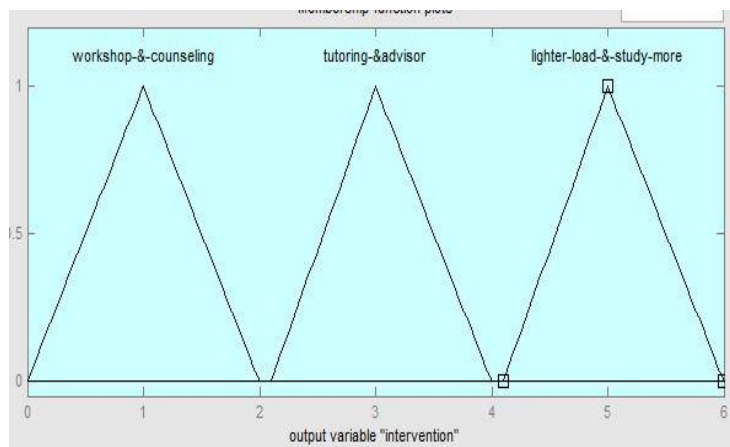

**Figure 5:** Graph of the Output Variable, Intervention

Along with the input variables are the rules. Rules are necessary to connect the linguistic variables to come up with the behavioral problem of students that include poor academic performance, tardiness and absenteeism. The rules are also given by the expert of the guidance counselors. Table 5 presents the rules.

**Table 5:** Rules for the Intervention of Student Behaviors

| Poor Academic Performance | Tardiness | Absenteeism | Intervention |
|---|---|---|---|
| **Low** | Low | Low | Workshop & Counseling |
| **Low** | Medium | Low | Workshop & Counseling |
| **Low** | High | Low | Workshop & Counseling |
| **Low** | Medium | Medium | Tutoring & Advisor |
| **Low** | High | Medium | Tutoring & Advisor |
| **Low** | High | High | Lighter load & Study more |
| **Medium** | Low | Low | Tutoring & Advisor |
| **Medium** | Low | Medium | Tutoring & Advisor |
| **Medium** | Medium | Medium | Lighter load & Study more |
| **Medium** | High | Medium | Tutoring & Advisor |
| **Medium** | Medium | High | Lighter load & Study more |
| **High** | Low | Low | Lighter load & Study more |
| **High** | Low | Medium | Lighter load & Study more |
| **High** | Medium | Low | Lighter load & Study more |
| **High** | Medium | Medium | Lighter load & Study more |
| **High** | High | High | Lighter load & Study more |

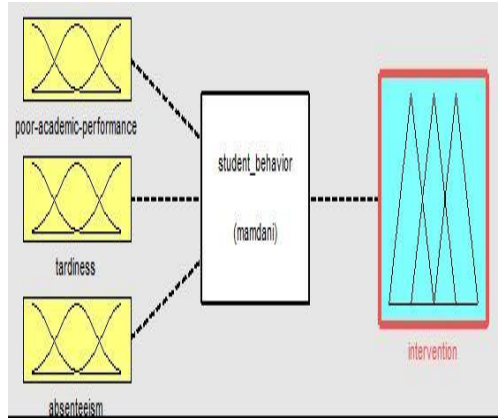

**Figure 6:** Intervention on Students' Behavior Fuzzy Inference System

## 4. Results and Discussion

To visualize the mapping from the student's behavior that include poor academic performance, tardiness and misconduct and the student's intervention, a Matlab software was used. Interventions include workshop and counseling, tutoring and advisor and lighter load and study more.

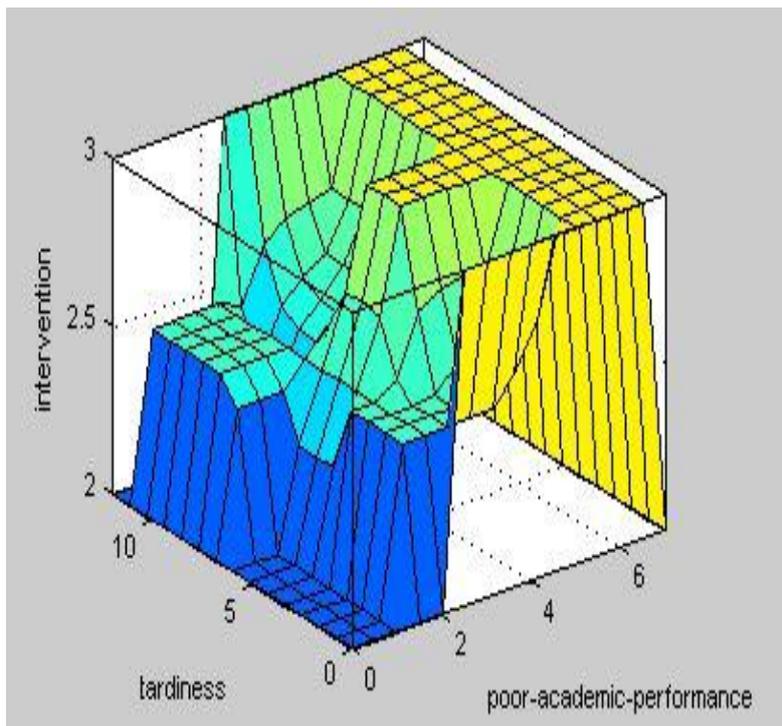

**Figure 7:** Three Dimensional Plot of the Rules for the Intervention on Students' Behavior

The three-dimensional graph presented in Figure 7 was noticeable that most of the students are referred on poor academic performance, followed by absenteeism then tardiness. For experimentation purposes, table 6 contained data obtained from the simulation in which several intervention are given.

**Table 6:** Data Obtained from Simulation

| Poor Academic Performance | Tardiness | Absenteeism | Intervention |
|---|---|---|---|
| **1** | 1 | 2 | Workshop & Counseling |
| **0** | 3 | 3 | Tutoring & Advisor |
| **3** | 5 | 5 | Lighter load & Study more |
| **5** | 7 | 2 | Lighter load & Study more |
| **2** | 6 | 6 | Lighter load & Study more |
| **7** | 5 | 4 | Tutoring & Advisor |
| **6** | 4 | 3 | Lighter load & Study more |
| **4** | 9 | 5 | Lighter load & Study more |
| **2** | 8 | 2 | Workshop & Counseling |
| **4** | 2 | 1 | Lighter load & Study more |

**Table 7:** The Intervention and its Frequency

| Intervention | Frequency |
|---|---|
| Workshop & Counseling | 2 |
| Tutoring & Advisor | 2 |
| Lighter load & Study more | 6 |

## 5. Conclusion and Recommendations

The visualized the academic behavior and mental ability test of all referred students of each college in Misamis University. The study successfully designs the data-driven decision support system for student's behavior.

The study gave the detailed design processes of identifying student's behavior that serves as an input to the management in designing interventions and making decisions. With the conduct of the study, the existing information in the Guidance and Testing Center are fully utilized, managed and analyzed. This contributed to the convenient procedure of deriving decision to help student's behavior problems. Thus, implementing decision support system help the management to make good decision making in terms of making interventions to the different behavioral problems of students.

The researcher recommended on the utilization of the designed algorithms by other higher education institutions who wish to determine the different behavioral problems pattern of their students.

**Declarations**

*Source of Funding*

This research did not receive any specific grant from funding agencies in the public, commercial, or not-for-profit sectors.

*Competing Interests Statement*

The authors declare no competing financial, professional and personal interests.

*Consent for publication*

We declare that we consented for the publication of this research work.

**References**

Ada, Ş., & Ghaffarzadeh, M. (2015). Decision making based on management information system and decision support system. European researcher. Series A, (4), 260-269.

Aiello, G., Giovino, I., Vallone, M., Catania, P., & Argento, A. (2018). A decision support system based on multisensor data fusion for sustainable greenhouse management. Journal of Cleaner Production, 172, 4057-4065.

Alshamrani, O. S., & Alshibani, A. (2020). Automated decision support system for selecting the envelope and structural systems for educational facilities. Building and Environment, 181, 106993.

Bellows, J., Patel, S., & Young, S. S. (2014). Use of IndiGO individualized clinical guidelines in primary care. Journal of the American Medical Informatics Association, 21(3), 432-437.

Belkadi, F., Dhuieb, M. A., Aguado, J. V., Laroche, F., Bernard, A., & Chinesta, F. (2020). Intelligent assistant system as a context-aware decision-making support for the workers of the future. Computers & Industrial Engineering, 139, 105732.

Bonczek, R. H., Holsapple, C. W., & Whinston, A. B. (2014). Foundations of decision support systems. Academic Press.

Brown-Liburd, H., Issa, H., & Lombardi, D. (2015). Behavioral implications of Big Data's impact on audit judgment and decision making and future research directions. Accounting Horizons, 29(2), 451-468.

Brynjolfsson, E., & McElheran, K. (2019). Data in Action: Data-Driven Decision Making and Predictive Analytics in US Manufacturing. Available at SSRN 3422397.

Datnow, A., & Hubbard, L. (2016). Teacher capacity for and beliefs about data-driven decision making: A literature review of international research. Journal of Educational Change, 17(1), 7-28.

Kombo, H. M. (2018). School Management Decisions and Students Academic Performance: A case of Kizimkazi and Dimbani Secondary Schools in Unguja, Tanzania (Doctoral dissertation, Mzumbe University).

Krishnaiyer, K., & Chen, F. F. (2017). Web-based visual decision support system (WVDSS) for letter shop. Robotics and Computer-Integrated Manufacturing, 43, 148-154.

Masmiquel, J. L., Ferrer, A. C., & Fonseca, D. (2017). School performance analysis from a scholastic learning process. Journal of Information Technology Research (JITR), 10(1), 1-14.

Mandinach, E., & Gummer, E. (2015). Data-driven decision making: Components of the enculturation of data use in education. Teachers College Record, 117(4), 1-8.

Marinakis, V., Doukas, H., Tsapelas, J., Mouzakitis, S., Sicilia, Á., Madrazo, L., & Sgouridis, S. (2018). From big data to smart energy services: An application for intelligent energy management. Future Gen. Computer Systems.

Miguéis, V. L., Freitas, A., Garcia, P. J., & Silva, A. (2018). Early segmentation of students according to their academic performance: A predictive modelling approach. Decision Support Systems, 115, 36-51.

Mohamed, A., Najafabadi, M. K., Wah, Y. B., Zaman, E. A. K., & Maskat, R. (2020). The state of the art and taxonomy of big data analytics: view from new big data framework. Artificial Intelligence Review, 53(2), 989-1037.

Norman, S. B., Rosen, J., Himmerich, S., Myers, U. S., Davis, B., Browne, K. C., & Piland, N. (2015). Student Veteran perceptions of facilitators and barriers to achieving academic goals. Journal of Rehabilitation Research & Development, 52(6).

Omidipoor, M., Jelokhani-Niaraki, M., Moeinmehr, A., Sadeghi-Niaraki, A., & Choi, S. M. (2019). A GIS-based decision support system for facilitating participatory urban renewal process. Land Use Policy, 88, 104150.

Owens, T. L. (2017). Higher education in the sustainable development goals framework. European Journal of Education, 52(4), 414-420.

Price, R., & Shanks, G. (2016). A semiotic information quality framework: development and comparative analysis. In Enacting Research Methods in Information Systems (pp. 219-250). Palgrave Macmillan, Cham.